\documentclass[letter]{aa} 
\usepackage{graphicx}
\usepackage{txfonts}
\begin{document} 
   \title{Herschel\footnote{Herschel is an ESA space observatory with science instruments provided by European-led Principal Investigator consortia and with important participation from NASA.} PEP/HerMES: On the redshift evolution (0 $\le$ z $\le$ 4) of dust attenuation and of the total (UV+IR) star formation rate density}

   \titlerunning{0 $\le$ z $\le$ 4 redshift evolution of the UV+IR SFRD and dust attenuation}
   
   \authorrunning{Burgarella, et al.}

   \subtitle{}

\author{D.~Burgarella\inst{1}
\and V.~Buat\inst{1}
\and C.~Gruppioni\inst{2}
\and O.~Cucciati\inst{2}
\and S.~Heinis\inst{1}
\and S.~Berta\inst{3}
\and M.~B{\'e}thermin\inst{4}
\and J.~Bock\inst{5,6}
\and A.~Cooray\inst{7,5}
\and J.S.~Dunlop\inst{8}
\and D.~Farrah\inst{9,10}
\and A.~Franceschini\inst{11}
\and E.~Le Floc'h\inst{4}
\and D.~Lutz\inst{3}
\and B.~Magnelli\inst{3}
\and R.~Nordon\inst{3}
\and S.J.~Oliver\inst{9}
\and M.J.~Page\inst{12}
\and P.~Popesso\inst{3}
\and F.~Pozzi\inst{13}
\and L.~Riguccini\inst{4}
\and M.~Vaccari\inst{11,14}
\and M.~Viero\inst{5}}
\institute{Aix-Marseille Universit\'e, CNRS, LAM (Laboratoire d'Astrophysique de Marseille) UMR7326, 13388, France\\
\email{denis.burgarella@oamp.fr}
\and INAF-Osservatorio Astronomico di Bologna, via Ranzani 1, I-40127 Bologna, Italy
\and Max-Planck-Institut f\"ur Extraterrestrische Physik (MPE), Postfach 1312, 85741, Garching, Germany
\and Laboratoire AIM-Paris-Saclay, CEA/DSM/Irfu - CNRS - Universit\'e Paris Diderot, CE-Saclay, pt courrier 131, F-91191 Gif-sur-Yvette, France
\and California Institute of Technology, 1200 E. California Blvd., Pasadena, CA 91125, USA
\and Jet Propulsion Laboratory, 4800 Oak Grove Drive, Pasadena, CA 91109, USA
\and Dept. of Physics \& Astronomy, University of California, Irvine, CA 92697, USA
\and Institute for Astronomy, University of Edinburgh, Royal Observatory, Blackford Hill, Edinburgh EH9 3HJ, UK
\and Astronomy Centre, Dept. of Physics \& Astronomy, University of Sussex, Brighton BN1 9QH, UK
\and Department of Physics, Virginia Tech, Blacksburg, VA 24061, USA
\and Dipartimento di Fisica e Astronomia, Universit\`{a} di Padova, vicolo Osservatorio, 3, 35122 Padova, Italy
\and Mullard Space Science Laboratory, University College London, Holmbury St. Mary, Dorking, Surrey RH5 6NT, UK
\and INAF-Osservatorio Astronomico di Roma, via di Frascati 33, 00040 Monte Porzio Catone, Italy
\and Astrophysics Group, Physics Department, University of the Western Cape, Private Bag X17, 7535, Bellville, Cape Town, South Africa}

   \date{Received April 5, 2013; accepted April 30, 2013 }

 
  \abstract
   {Using new homogeneous Luminosity Functions (LFs) in the FUV (VVDS) and in the FIR {\sl Herschel}/PEP and {\sl Herschel}/HerMES, we study the evolution of the dust attenuation with redshift. With this information in hand, we are able to estimate the redshift evolution of the total (FUV + FIR) star formation rate density (${\rm SFRD_{TOT}}$). By integrating ${\rm SFRD_{TOT}}$, we follow the mass building and analyze the redshift evolution of the stellar mass density (SMD). }
{ This letter aims at providing a complete view of star formation from the local universe to $z \sim 4$ and, using assumptions on earlier star formation history, compares this evolution to what was known before in an attempt to draw a homogeneous picture of the global evolution of star formation in galaxies.
}
{ }
{The main conclusions of this letter are: 1) the dust attenuation $A_{FUV}$ is found to increase from $z = 0$ to $z \sim 1.2$ and then starts to decrease up to our last data point at $z = 3.6$; 2) the estimated SFRD confirms published results up to $z \sim 2$. At $z > 2$, we observe either a plateau or a small increase up to $z \sim 3$ and then a likely decrease up to $z = 3.6$; 3) the peak of ${\rm A_{FUV}}$ is delayed with respect to the plateau of ${\rm SFRD_{TOT}}$ and a likely origin might be found in the evolution of the bright ends of the FUV and FIR LFs; 4) using assumptions (namely exponential rise and linear rise with time) for the evolution of the star formation density from $z = 3.6$ to $z_{form} = 10$, we integrate ${\rm SFRD_{TOT}}$ and find a good agreement with the published SMDs.}

   \keywords{ Galaxies: starburst --   Ultraviolet: galaxies --  Infrared: galaxies -- Galaxies: high-redshift -- Cosmology: early universe}

   \maketitle
%

\section{Introduction}

One of the major objectives in astrophysics during the last 15 years or so has been to follow the cosmic star formation rate density (SFRD) at earlier and earlier epochs. But, whenever optical data are used, one must apply a dust correction to luminosity densities (LDs) and a calibration into SFRDs (with their associated uncertainties) to obtain a relevant estimate. Knowing how the dust attenuation evolves in redshift is therefore mandatory if one wishes to study the redshift evolution of the SFRD.

For instance, \cite{Takeuchi05} estimate the cosmic evolution of the SFRD from far-ultraviolet (FUV) and far-infrared (FIR = bolometric IR). They find an increase of the fraction of hidden SFR from 56\% locally to 84\% at $z = 1$. The LDs show a significant evolution as the FIR LD evolves faster than the FUV. Their ratio $\rho_{FIR}/\rho_{FUV}$ increases from $\sim$  4 (A$_{FUV} \sim 1.3$ mag) locally to $\sim$ 15 (A$_{FUV} \sim 2.3$ mag) at $z = 1$. \cite{Cucciati12} used the VIMOS-{\sl VLT} Deep Survey to show, from the FUV only that the mean dust attenuation A$_{FUV}$ is in agreement with \cite{Takeuchi05} over the range $0 < z < 1$. Then it remains at the same level up to $z \sim 2$, and declines to $\sim$ 1 mag at $z \sim 4$. 

In this letter, we use the FUV luminosity functions (LFs) published by \cite{Cucciati12} from the VLT along with the FIR LFs from {\sl Herschel}-PACS+SPIRE data\footnote{From two {\sl Herschel} Large Programmes: PACS Evolutionary Probe (\cite{Lutz11}) and the {\sl Herschel} Multi-tiered Extragalactic Survey (HerMES, \cite{Oliver12})} of a PACS selected sample from \cite{Gruppioni13} to constrain the redshift evolution of  $\log_{10} (L_{FIR} / L_{FUV}$) (aka $IRX$) up to $z \sim 4$ for the first time directly using FIR data. With this information in hand, we can estimate the redshift evolution of $\rho_{FIR}/\rho_{FUV}$ as well as $\rho_{TOT} = \rho_{FIR} + \rho_{FUV}$. Finally, by integrating $\rho_{TOT}$, we estimate the cosmic evolution of stellar mass density (SMD) with redshift.

Throughout this paper we adopt a $\Lambda$CDM cosmology with ($H_0, \Omega_m, \Omega_\Lambda$) = (70, 0.3, 0.7), where $H_0$ is in kms$^{-1}$Mpc$^{-1}$. All SFR and stellar masses presented assume, or have been converted to, a Salpeter IMF.

\section{Luminosity Functions}

Our analysis at $z\sim0$ is based on the FUV LF from \cite{Wyder05} and the FIR LF from \cite{Takeuchi05}, and for $0 < z < 4$ on the FUV LF from \cite{Cucciati12} and the FIR LF from \cite{Gruppioni13}. In the FIR and at $z > 0$, the sample is selected in the PACS bands but uses the full {\sl Herschel}-PACS + {\sl Herschel}-SPIRE SED data. The PACS selection means that we can miss sources towards the upper end of the redshift range. The LFs are evaluated from homogeneous datasets in the FUV and the FIR. This minimizes biases and keeps the same reference indicator throughout cosmic times with a simple well-defined and controlled selection function. This is one of the strengths of this work. The FUV LFs are not corrected for dust attenuation. We define the LFs as a number density of galaxies with luminosity in logarithmic intervals, [${\rm log_{10} L, log_{10} L + d log_{10} L}$], where ${\rm \Phi (L) = dn / d log_{10} L}$ and the luminosity is defined as ${\rm L \equiv \nu L_\nu}$. FIR luminosities are defined as: ${\rm \Phi(L) = \Phi_\star (\frac{L}{L_\star})^{1-\alpha} exp(-\frac{1}{2 \sigma^2} [\log_{10} (1 + \frac{L}{L_\star} )]^2)}$.

Observed uncertainties from \cite{Cucciati12} and from \cite{Gruppioni13} are used whenever available. However, some of the Schechter parameters are fixed when the LFs are derived, namely $\alpha$ for the FUV LFs and $\alpha$ plus $\sigma$ for the FIR LFs. Both in FUV and in FIR, we assume uncertainties of 10\% up to $z = 1$, 20\% up to $z = 2$ and 40\% beyond for these fixed parameters. This level of uncertainty is similar to previous works in FUV by e.g. \cite{Oesch10, VandenBurg10} and in FIR by \cite{Casey12}. We propagate uncertainties by simulating 2000 realizations drawn from 1-$\sigma$ Gaussian distributions for each parameter with known uncertainties and from a flat distribution (i.e. equiprobability) for the fixed ones. We assume that all fixed values are equiprobable given the weak observational constraints. Finally, we interpolate the FUV and FIR Schechter parameters on the same redshift grid between $z = 0$ and $z = 3.6$.

\begin{figure*}
   \centering
   \includegraphics[width=\hsize]{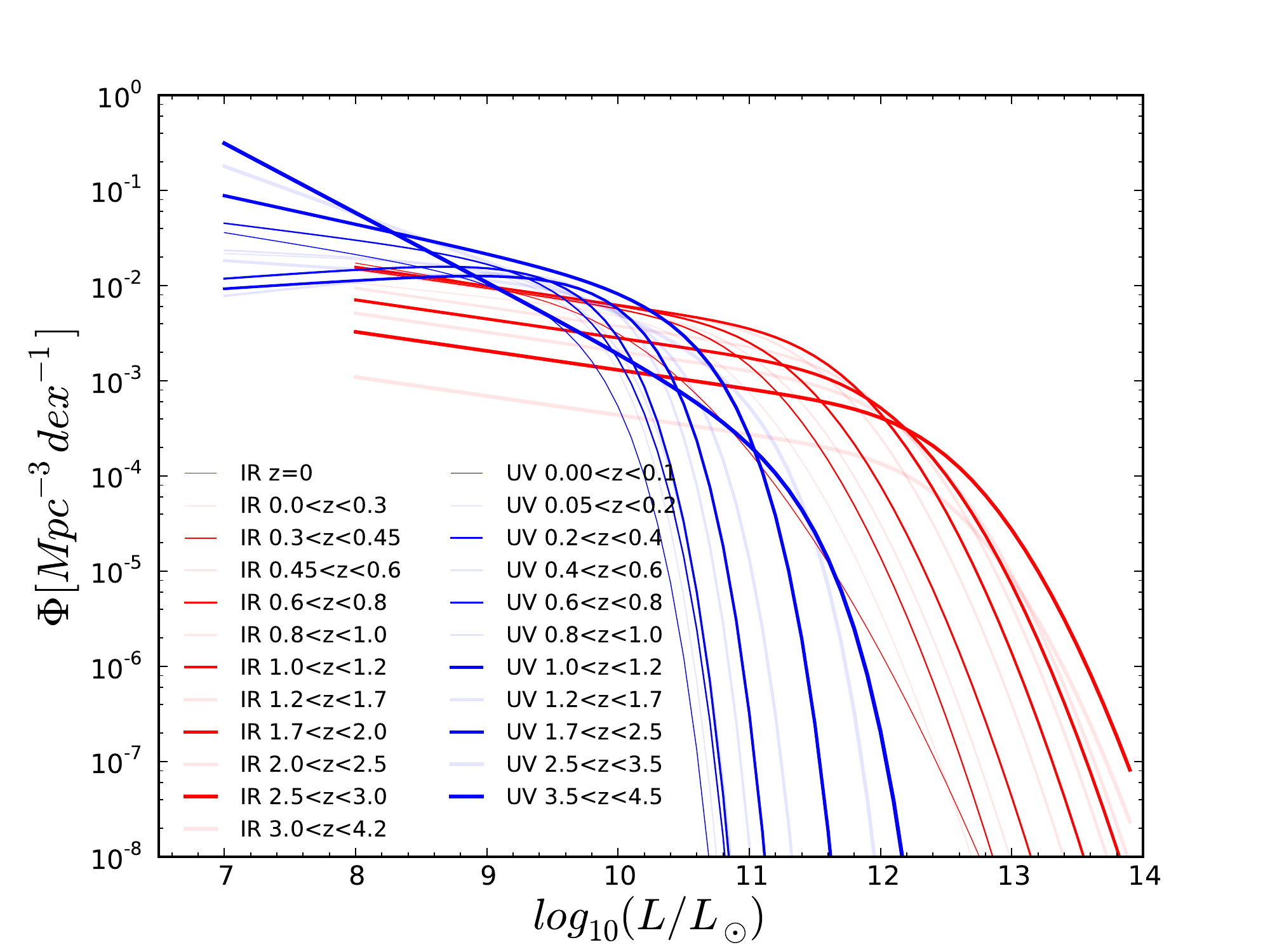}
      \caption{Redshift evolution of the FIR from \cite{Gruppioni13} (red) and FUV from \cite{Cucciati12} (blue) LFs. Note that the FUV LFs are uncorrected for dust attenuation. The LFs at every other redshift are plotted boldly. The others are fainter to lighten the figure. The LFs are plotted within the limits of integration.
              }
         \label{FigLFs}
   \end{figure*}

Tab.~\ref{TabLFs} and Fig.~\ref{FigLFs} show the redshift variation of the LFs in FIR. The known difference in the FIR and FUV LFs  \cite{Takeuchi05} are clearly illustrated here: bright FIR galaxies are more numerous than bright FUV galaxies at $log_{10} (L [L_\odot]) > 10$. In FUV, except in the highest redshift bins, L$^\star$ and ${\Phi^\star}$ remain approximately constant while the faint-end slope evolves. The FIR faint end slope is not observationally constrained at high z, and \cite{Cucciati12} fix it to $\alpha = 1.2$. However, L$^\star$ and ${\Phi^\star}$ are allowed to change with redshift. These different evolutions of the FUV and FIR LFs are reflected in Fig.~\ref{FigLFs} and explain the evolution of the cosmic SFRD and dust attenuation.

\begin{table*}
\caption{\label{t1}Schechter parameter for FUV and FIR luminosity functions.}
\centering
\resizebox{\textwidth}{!}{ 
\begin{tabular}{cccccc}
\hline\hline
Redshift range & ${\rm M^\star}$ or ${\rm L^\star}$\tablefootmark{a} & ${\rm \Phi_\star}$ & $\alpha$ & $\sigma$\tablefootmark{b} \\
\hline
&  &  FUV luminosity functions  &  &  \\
\hline 
$0.0 < z < 0.1$\tablefootmark{c}     & -18.04 $\pm$ 0.11 & -2.370  $\pm$ 0.06 & -1.22 $\pm$  0.07 & --- \\
$0.05 < z < 0.2$\tablefootmark{d}  & -18.12 $\pm$ 0.00 & -2.155  $\pm$ 0.03 &  -1.05 $\pm$ 0.04 & ---  \\
$0.2 < z < 0.4$\tablefootmark{d}    & -18.3 $\pm$    0.20 & -2.161 $\pm$  0.06 & -1.17 $\pm$  0.05 & --- \\
$0.4 < z < 0.6$\tablefootmark{d}    & -18.4  $\pm$   0.10 & -2.180 $\pm$  0.06 & -1.07 $\pm$  0.07 & --- \\
$0.6 < z < 0.8$\tablefootmark{d}    & -18.3 $\pm$    0.10 & -2.021 $\pm$  0.05 & -0.90 $\pm$  0.08 & --- \\
$0.8 < z < 1.0$\tablefootmark{d}    & -18.7  $\pm$   0.10 & -2.045 $\pm$  0.05 & -0.85 $\pm$  0.10 & --- \\
$1.0 < z < 1.2$\tablefootmark{d}    & -19.0  $\pm$   0.20 & -2.129  $\pm$ 0.07 & -0.91 $\pm$  0.16 & --- \\
$1.2 < z < 1.7$\tablefootmark{d}    & -19.6  $\pm$   0.20 & -2.387  $\pm$ 0.10 & -1.09 $\pm$  0.23 & --- \\
$1.7 < z < 2.5$\tablefootmark{d}    & -20.4 $\pm$    0.10 & -2.472  $\pm$ 0.03 & -1.30 $\pm$ -0.26 & --- \\
$2.5 < z < 3.5$\tablefootmark{d}    & -21.4 $\pm$    0.10 & -3.066  $\pm$ 0.03 & -1.50 $\pm$ -0.60 & --- \\
$3.5 < z < 4.5$\tablefootmark{d}    & -22.2 $\pm$    0.20 & -3.959  $\pm$ 0.04 & -1.73 $\pm$ -0.69 & --- \\
\hline\hline
&  &  FIR luminosity functions  &  &  \\
\hline 
$z = 0$\tablefootmark{e}                &   9.25 $\pm$ 0.00 & -2.051 $\pm$ 0.00 & 1.23 $\pm$ 0.00 & 0.72 $\pm$ 0.00 \\
$0.0 < z < 0.3$\tablefootmark{f}    & 10.12 $\pm$ 0.16 & -2.29   $\pm$ 0.06 & 1.15 $\pm$ 0.05 & 0.52 $\pm$ 0.05 \\
$0.3 < z < 0.45$\tablefootmark{f}  & 10.41$\pm$  0.03 & -2.31   $\pm$ 0.03 & 1.2  $\pm$ -0.12 & 0.5  $\pm$ -0.05 \\
$0.45 < z < 0.6$\tablefootmark{f}  & 10.55$\pm$  0.03 & -2.35   $\pm$ 0.05 & 1.2  $\pm$ -0.12 & 0.5  $\pm$ -0.05 \\
$0.6 < z < 0.8$\tablefootmark{f}    & 10.71 $\pm$ 0.03 & -2.35   $\pm$ 0.06 & 1.2  $\pm$ -0.12 & 0.5  $\pm$ -0.05 \\
$0.8 < z < 1.0$\tablefootmark{f}    & 10.97 $\pm$ 0.04 & -2.40   $\pm$ 0.05 & 1.2  $\pm$ -0.12 & 0.5  $\pm$ -0.05 \\
$1.0 < z < 1.2$\tablefootmark{f}    & 11.13 $\pm$ 0.04 & -2.40   $\pm$ 0.05 & 1.2  $\pm$ -0.24 & 0.5  $\pm$ -0.10 \\
$1.2 < z < 1.7$\tablefootmark{f}    & 11.37 $\pm$ 0.03 & -2.70   $\pm$ 0.04 & 1.2  $\pm$ -0.24 & 0.5  $\pm$ -0.10 \\
$1.7 < z < 2.0$\tablefootmark{f}    & 11.50 $\pm$ 0.03 & -2.85   $\pm$ 0.03 & 1.2  $\pm$ -0.24 & 0.5  $\pm$ -0.10 \\
$2.0 < z < 2.5$\tablefootmark{f}    & 11.60 $\pm$ 0.03 & -3.01   $\pm$ 0.11 & 1.2  $\pm$ -0.48 & 0.5  $\pm$ -0.20 \\
$2.5 < z < 3.0$\tablefootmark{f}    & 11.92 $\pm$ 0.08 & -3.27   $\pm$ 0.18 & 1.2  $\pm$ -0.48 & 0.5  $\pm$ -0.20 \\
$3.0 < z < 4.2$\tablefootmark{f}    & 11.90 $\pm$ 0.16 & -3.74   $\pm$ 0.12 & 1.2  $\pm$ -0.48 & 0.5  $\pm$ -0.20 \\

\hline
\end{tabular}}
\tablefoot{The top panel lists the FUV LFs and the bottom panel lists the FIR LFs. For all the parameters with an uncertainty set to 0.00, we assumed 20\% or error.
\tablefoottext{a}{${\rm L^\star}$ [L$_\odot$] for FIR LFs or ${\rm M^\star}$ [AB mag] for FUV LFs}
\tablefoottext{b}{$\sigma$ only needed for FIR LFs}
\tablefoottext{c}{from \cite{Wyder05}}
\tablefoottext{d}{from \cite{Takeuchi05}}
\tablefoottext{e}{\cite{Cucciati12}}
\tablefoottext{f}{\cite{Gruppioni13}}
}
 \label{TabLFs}
\end{table*}

\section{Dust Attenuation traced by the FIR to FUV LD ratio}

\begin{figure*}
   \centering
   \includegraphics[width=\hsize]{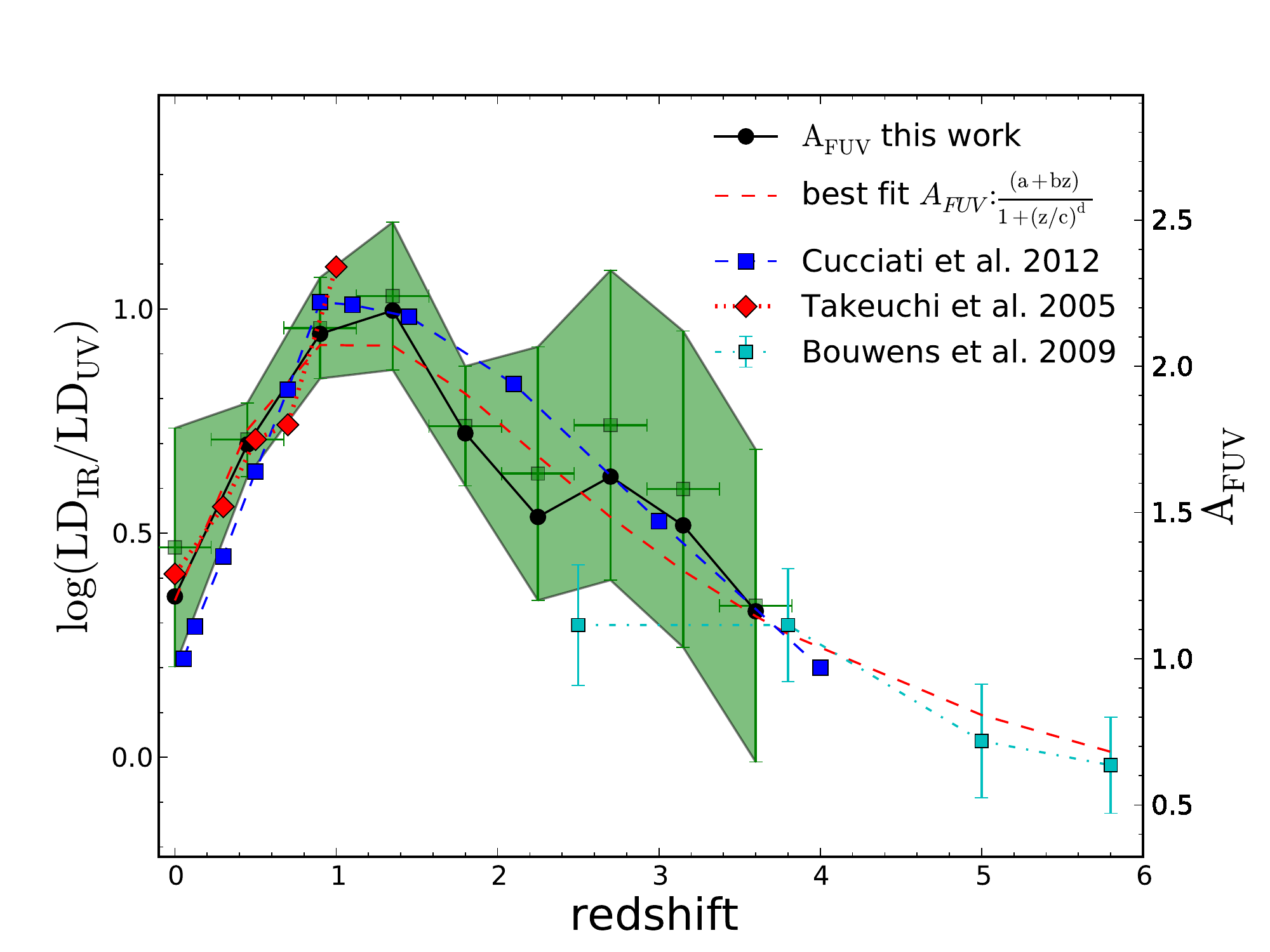}
      \caption{Left axis: ratio of the FIR to FUV LDs ($IRX$). Right axis: FUV dust attenuation (A$_{FUV}$). The red dotted line with red diamonds is from Takeuchi, Buat \& Burgarella (2005). The green filled area and green dots are the associated uncertainties estimated through bootstraping with 2 000 drawings. Black dots are the values directly computed from the LFs. At $z = 3.6$, A$_{FUV}$ reaches about the same value as at $z = 0$. \cite{Takeuchi05} (red diamonds) used an approach identical to ours while a SED analysis (no FIR data) is performed in \cite{Cucciati12} (blue boxes). \cite{Bouwens09} are estimates based on the UV slope $\beta$. The limiting FUV luminosity is $10^{7} L_\odot$ or $1.65\times10^{-4} L^\star_{z=3}$. The best fit is given by ${\rm A_{FUV} (z) =\frac{(a+bz)}{1+(z/c)^d}}$ with $a = 1.20$, $b = 1.50$, $c = 1.77$, and $d = 2.19$.
              }
         \label{FigAfuv}
   \end{figure*}

Fig.~\ref{FigAfuv} presents the dust attenuation in the FUV vs. z and the ratio of the FIR to FUV LDs integrated in the range $log_{10} (L [L_\odot$]) = [7, 14] in the FUV (i.e. L$^{min}_{FUV}$ =  $1.65\times10^{-4} L^\star_{z=3}$, \cite{Bouwens09}) and [8, 14] in the FIR. The FUV dust attenuation is estimated from the $IRX$ and converted to A$_{FUV}$ using \cite{Burgarella05}\footnote{The conversion from $IRX$ to A$_{FUV}$ from \cite{Burgarella05} is valid at $\log_{10}$ (L$_{FIR}$/L$_{FUV}$)$>$-1.2: A$_{FUV}$ = -0.028 [$\log_{10}$ L$_{FIR}$/L$_{FUV}$]$^3$ + 0.392 [$\log_{10}$ L$_{FIR}$/L$_{FUV}$]$^2$ + 1.094 [$\log_{10}$ L$_{FIR}$/L$_{FUV}$] + 0.546}. The redshift evolution of A$_{FUV}$ is in agreement with \cite{Cucciati12}. Note that \cite{Cucciati12} estimated A$_{FUV}$ through an analysis of individual SEDs up to $\lambda_{obs} = 2.2 \mu$m (Ks-band). Fig.~\ref{FigAfuv} suggests the presence of a local minimum at $z \sim 2$ that might be due to UV-faint galaxies (see Fig.~7 in \cite{Cucciati12}) that are responsible for a peak observed in the FUV LD and not observed in the FIR. Since the fields observed in FUV and in FIR are not the same ones, another origin might be found in cosmic variance. The bottomline is that the existence of this trough in A$_{FUV}$ seems dubious. Finally, higher redshift $A_{FUV}$ from the UV slope, $\beta$, suggest a continuous decline at least up to $z = 6$ (\cite{Bouwens09}). 

We conclude that the cosmic dust attenuation A$_{FUV}$ reaches an absolute maximum at $z \sim 1.2$ followed by a global decline to $z = 3.6$ where it reaches about the same level measured at $z = 0$. 

%



\begin{figure*}
   \centering
   \includegraphics[width=\hsize]{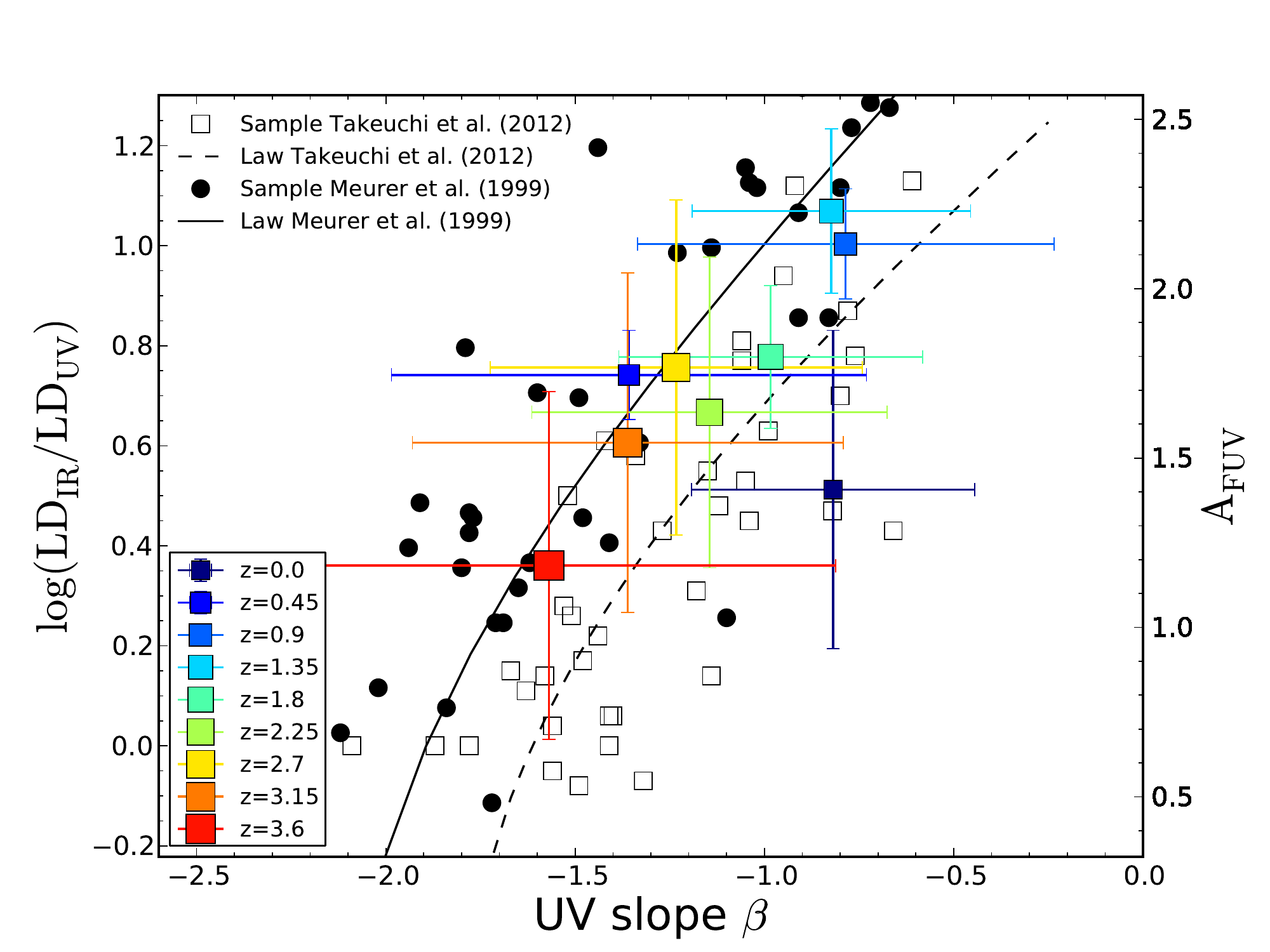}
      \caption{Dust attenuation vs. redshift. The boxes are color-coded according to redshift. Note that the x-axis bars correspond to the dispersion in UV slope while the y-axis are evaluated from the uncertainties in the LFs. The black dots are the original data points from \cite{Meurer99} and the black curve is \cite{Meurer99}'s law. The black dashed line and grey boxes are the update (using the same apertures in FUV and in FIR) from \cite{Takeuchi12}. Strictly speaking, our points and those from \cite{Meurer99} are not comparable because we use volume-corrected LFs and not individual galaxies as done in \cite{Meurer99}. In the diagram, they show an almost continuous decrease with increasing redshift and lie in between the \cite{Meurer99} and \cite{Takeuchi12} laws. It is important to stress that the dust attenuation A$_{FUV}$ are estimated from the $IRX$ and not from the UV slope $\beta$.
              }
         \label{FigIRXbeta}
   \end{figure*}

The $\beta$ method is popular because estimates the total SFR from the FUV only. This is most useful at high redshifts where the samples are UV-selected (\cite{Burgarella11, Bouwens12, Heinis13}). We propose to follow the redshift evolution of the cosmic volume-averaged points in the $IRX-\beta$ diagram (Fig.~\ref{FigIRXbeta}) to constrain models. However, we must caution that the values plotted in Fig.~\ref{FigIRXbeta} cannot be directly compared to galaxies. The x-axis is calculated from the averaged rest-frame FUV - near-UV colors (\cite{Cortese06}). Horizontal error bars indicate the dispersion of the FUV slope. The $IRX$ is estimated from LFs and is therefore volume-corrected. Vertical error bars are uncertainties. This $IRX-\beta$ plot can be interpreted as the location of a comoving volume as a function of redshift. From $z \sim 1$ to $z \sim 4$, the points evolve downwards parallel to the original \cite{Meurer99} law and the update by \cite{Takeuchi12}.


\section{The total FUV+FIR Star Formation Density and Stellar Mass Density}


The calibration from LD to SFRD is problematic (\cite{Kobayashi13}) in the FUV and also in the FIR (\cite{Kennicutt98, Schaerer13}). In agreement with \cite{Casey12}, we use \cite{Kennicutt98} calibrations and assume a Salpeter initial mass function (IMF) to allow a better comparison with other published SFRDs. Note that the AGN contribution to the FIR LDs has been estimated in FIR using a SED fitting and subtracted off i.e. the presented FIR LD is due to star formation only. 

\begin{figure*}
   \centering
   \includegraphics[width=\hsize]{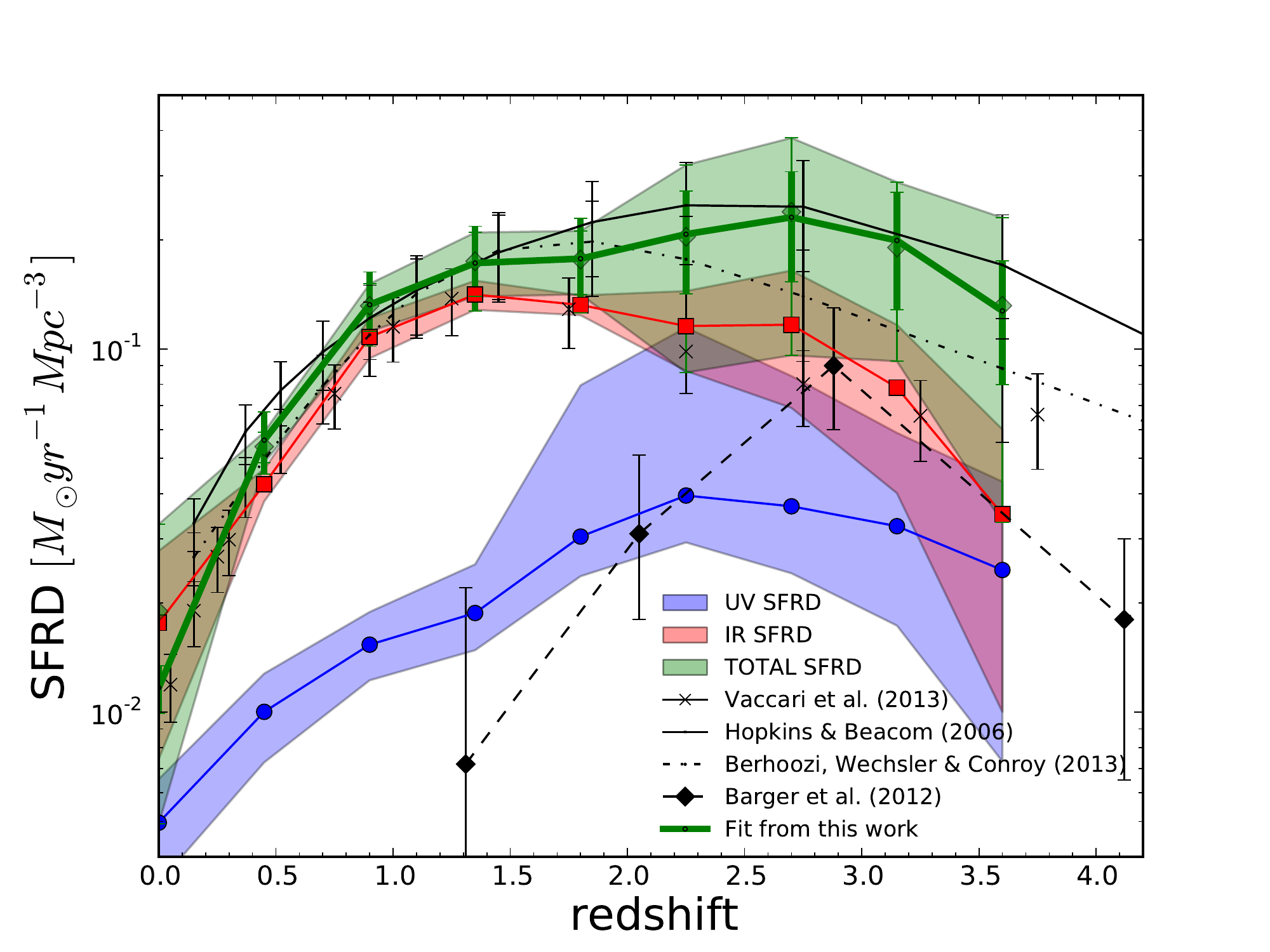}
      \caption{SFRD densities in the FUV (blue), in the FIR (red) and in total (i.e. FUV + FIR) in green (other colors are due to overlaps of the previous colors). The lines are the mean values while the lighter colors shows the uncertainties evaluated from the 2 000 runs as in Fig.~\ref{FigAfuv}. After the initial increase of the total SFRD from $z = 0$ to $z \sim 1.2$, it remains flat or slightly increases/decreases up to $z \sim 2.5 - 3.0$ followed by a decrease. Globally and over $0 < z \le 3.6$, the total average SFRD is slightly below  \cite{Hopkins06}'s and in agreement with \cite{Behroozi12} up to $z \sim 2.$. SFRD from \cite{Barger12} and preliminary results from {\sl Herschel}/SPIRE estimated by \cite{Vaccari13} are in agreement with these trends. Symbols and lines are explained in the plot.
              }
         \label{FigSFRD}
   \end{figure*}

Fig.~\ref{FigSFRD} suggests a flattening of the total SFRD up to $z \sim 3$ (\cite{Chary01, Perez-Gonzalez05, LeFloch05, Franceschini10}) where the UV data favour a peak followed by a decrease. Note that we could not rule out a small increase or decrease within the uncertainties. The plateau up to $z\sim2.5$ is seen by \cite{Rodighiero10} and \cite{Magnelli11} while the decrease of $\rho_{SFR}$ at $z\ge2.5$ is predicted by \cite{Bethermin12b}'s model based on evolution of mass function and sSFR estimated from LBGs. All in all, our total SFRD is in fair agreement with that of \cite{Hopkins06} in the same redshift range. However, discrepancies exist: our total SFRD is lower at $z < 1$ and is only marginally consistent but lower at $z > 3$. Also, note that PACS data are less sensitive at higher than at lower redshift since the rest-frame wavelength moves into the mid-IR. The preliminary FIR SFRD from \cite{Vaccari13} ({\sl Herschel}/SPIRE selection) is in excellent agreement over the $0 < z \le 2$ range but is slightly higher than that derived from PACS at $z > 3$. However, this is only a $\sim 2\sigma$ difference. \cite{Barger12} published a FIR SFRD based on SCUBA-2 data that is also in agreement with ours at $2 < z < 4$. We first tried to fit  ${\rm SFRD_{TOT}}$ with a one-peak analytical function (\cite{Hopkins06, Behroozi12}) but the results are not satisfactory. So, we combined two Gaussians: $$a_1 e^{\frac{-(z-z_1)^2}{2 \sigma_1^2}} + a_2 e^{\frac{-(z-z_2)^2}{2 \sigma_2^2}}$$ with $a_1 = 0.1261 \pm 0.0222$, $\sigma_1 = 0.5135 \pm 0.0704$, $z_1 = 1.1390 \pm 0.0959$ and $a_2 = 0.2294 \pm 0.0222$, $\sigma_2 = 0.8160 \pm 0.0964$, $z_2 = 2.7151 \pm 0.0839$. At higher redshifts, we made assumptions that are explained below.

The cosmic SFRD presents a (weak) maximum at $z \sim 2.5 - 3.0$ (i.e. between 2.6 - 2.1 Gyrs  resp.) while the dust attenuation presents a maximum at $z \sim 1.2$ (i.e. 5 Gyrs). We have tried to lock the faint end slope of the UV LF -1.2, to see how far out in redshift the peak of obscuration could potentially move and we do not detect any change, suggesting this effect is solid. We have no definite explanation for this delay of $\sim$ 2.7 Gyr. Type II supernovae start producing dust earlier than AGB stars (\cite{Valiante09}) but the difference in timescales is too short and only on the order of a few 10 Myr for the onset of dust formation. Dust grain destruction in the ISM might play a role (\cite{Dwek11}) but the efficiency of destruction is not well known and depends on the star formation history. These dust-related origins for the delayed maximum are not likely. The best explanation might be that this delay is related to a global move of galaxies in the [$\log_{10} (L_{FIR} / L_{FUV}$) vs. $\log_{10} (L_{FIR} + L_{FUV})$] diagram. \cite{Buat09} showed that galaxies evolve in redshift from $z = 0$ to $z = 2$ in this diagram, with high redshift sources having lower $IRX$ at given total luminosities. This change is likely related to systematic changes of the FIR SEDs themselves (\cite{Elbaz11, Nordon12}). This suggests that the shift might be caused by the relative importance of more luminous galaxies ($\log_{10} L_{FUV} [L_\odot] \ge 10$) in the FUV as z evolves.

\begin{table*}
\caption{\label{t2}Evolution of the FUV and FIR LDs, the mean dust attenuation and the cosmic SFRDs in FUV and FIR. ${\rm SFR_{Hidden}}$ corresponds to the percentage of SFRD in dust, ${\rm SFRD_{TOT}}$ is the total (i.e. FUV + FIR) SFRD and $\beta$ is the mean slope estimated for the galaxies in the FUV sample. Lower limits of integration are ${\rm L^{min}_{FUV}  = 10^7 L_\odot}$ and ${\rm L^{min}_{FIR}  = 10^8 L_\odot}$ and upper limits are ${\rm L^{max}_{UV}  = L^{max}_{FIR} = 10^{14} L_\odot}$.}
\centering
\resizebox{\textwidth}{!}{ %
\begin{tabular}{cccccccccc}
\hline\hline
${\rm z_{mean}}$  & ${\rm LD_{IR} / LD_{UV}}$ & ${\rm A_{FUV}}$\tablefootmark{a} & ${\rm SFRD_{FUV}}$\tablefootmark{b}  & ${\rm SFRD_{FIR}}$\tablefootmark{c} & ${\rm SFRD_{TOT}}$\tablefootmark{d} & ${\rm SFRD_{Hidden}}$\tablefootmark{e} & $\beta$\tablefootmark{f} \\
  & & [mag] & [${\rm 10^{-2} M_\odot yr^{-1}Mpc^{-3}}$] & [${\rm 10^{-2} M_\odot yr^{-1}Mpc^{-3}}$]  & [${\rm 10^{-2} M_\odot yr^{-1}Mpc^{-3}}$]  &  fraction &  \\
\hline
0.0 & 4.25 $\pm$ 3.25 & 1.38 $\pm$ 0.40 & 0.50 $^{0.16} _{0.16}$ & 1.76 $\pm$ 1.02 & 1.86 $\pm$ 0.99 & 0.69 $\pm$ 0.10 & -0.87 $\pm$ 0.37 \\ 
0.45 & 7.11 $\pm$ 1.29 & 1.75 $\pm$ 0.13 & 1.00 $^{0.28} _{0.28}$ & 4.24 $\pm$ 0.43 & 5.41 $\pm$ 0.54 & 0.82 $\pm$ 0.03 & -1.41 $\pm$ 0.63 \\ 
0.9 & 11.84 $\pm$ 2.75 & 2.14 $\pm$ 0.17 & 1.54 $^{0.35} _{0.31}$ & 10.80 $\pm$ 1.37 & 13.10 $\pm$ 1.93 & 0.88 $\pm$ 0.03 & -0.84 $\pm$ 0.55 \\ 
1.35 & 13.44 $\pm$ 4.68 & 2.24 $\pm$ 0.25 & 1.88 $^{0.68} _{0.39}$ & 14.13 $\pm$ 1.31 & 17.44 $\pm$ 3.39 & 0.89 $\pm$ 0.04 & -0.87 $\pm$ 0.37 \\ 
1.8 & 7.66 $\pm$ 2.40 & 1.80 $\pm$ 0.21 & 3.05 $^{4.90} _{0.68}$ & 13.22 $\pm$ 0.83 & 17.70 $\pm$ 3.55 & 0.82 $\pm$ 0.04 & -1.03 $\pm$ 0.40 \\ 
2.25 & 6.03 $\pm$ 4.97 & 1.63 $\pm$ 0.45 & 3.95 $^{7.51} _{1.02}$ & 11.58 $\pm$ 2.87 & 19.95 $\pm$ 11.73 & 0.74 $\pm$ 0.11 & -1.20 $\pm$ 0.47 \\ 
2.7 & 7.59 $\pm$ 6.99 & 1.79 $\pm$ 0.51 & 3.69 $^{4.74} _{1.28}$ & 11.67 $\pm$ 4.78 & 24.03 $\pm$ 14.85 & 0.75 $\pm$ 0.15 & -1.28 $\pm$ 0.49 \\ 
3.15 & 5.54 $\pm$ 5.59 & 1.56 $\pm$ 0.52 & 3.25 $^{2.63} _{1.52}$ & 7.83 $\pm$ 3.82 & 19.06 $\pm$ 10.28 & 0.68 $\pm$ 0.18 & -1.41 $\pm$ 0.57 \\ 
3.6 & 3.16 $\pm$ 3.55 & 1.19 $\pm$ 0.52 & 2.46 $^{1.85} _{1.73}$ & 3.51 $\pm$ 2.51 & 12.84 $\pm$ 9.68 & 0.53 $\pm$ 0.23 & -1.62 $\pm$ 0.76 \\ 
\hline
\end{tabular}}
\tablefoot{
\tablefoottext{a}{calibrated from ${\rm \log_{10} LD_{IR} / LD_{UV}}$ using \cite{Burgarella05}. Changing the limits of integration to $log_{10} (L [L_\odot$]) = [4, 14] (resp. [9, 14]) in FUV and [5, 14] (resp. [10, 14]) in FIR would change A$_{FUV}$ by $< 0.1$ (resp. $\sim -0.2$ ) at $z \sim 3-4$ and $< 0.05$ below $z < 2$.}
\tablefoottext{b}{calibrated from ${\rm LD_{UV}}$ using \cite{Kennicutt98}}
\tablefoottext{c}{calibrated from ${\rm LD_{IR}}$ using \cite{Kennicutt98}}
\tablefoottext{d}{computed as ${\rm SFRD_{UV}}$+${\rm SFRD_{IR}}$. Note that the values presented in this column are larger than ${\rm SFRD_{FUV}+ SFRD_{FUV}}$ from the two previous columns because it is the mean of the 2000 realizations estimated from the LFs which happen to be above the sum of the SFRDs estimated by \cite{Cucciati12} and \cite{Gruppioni13}. Changing the limits of integration to $log_{10} (L [L_\odot$]) = [4, 14] (resp. [9, 14]) in FUV and [5, 14] (resp. [10, 14]) in FIR would change ${\rm SFRD_{TOT}}$ by less than +8\% (resp. -8\%) at $z \le 2.7$ but by + 56\% and + 158\% at ${\rm z_{mean}}$ = 3.15 and ${\rm z_{mean}}$=3.6 (resp -15\% and -32\%).}
\tablefoottext{e}{computed as ${\rm SFRD_{UV}}$/${\rm SFRD_{IR}}$}
\tablefoottext{h}{This column presents the mean $\beta$ only for the objects detected in UV.}
}
 \label{TabData}
\end{table*}



%

\begin{figure*}
   \centering
   \includegraphics[width=\hsize]{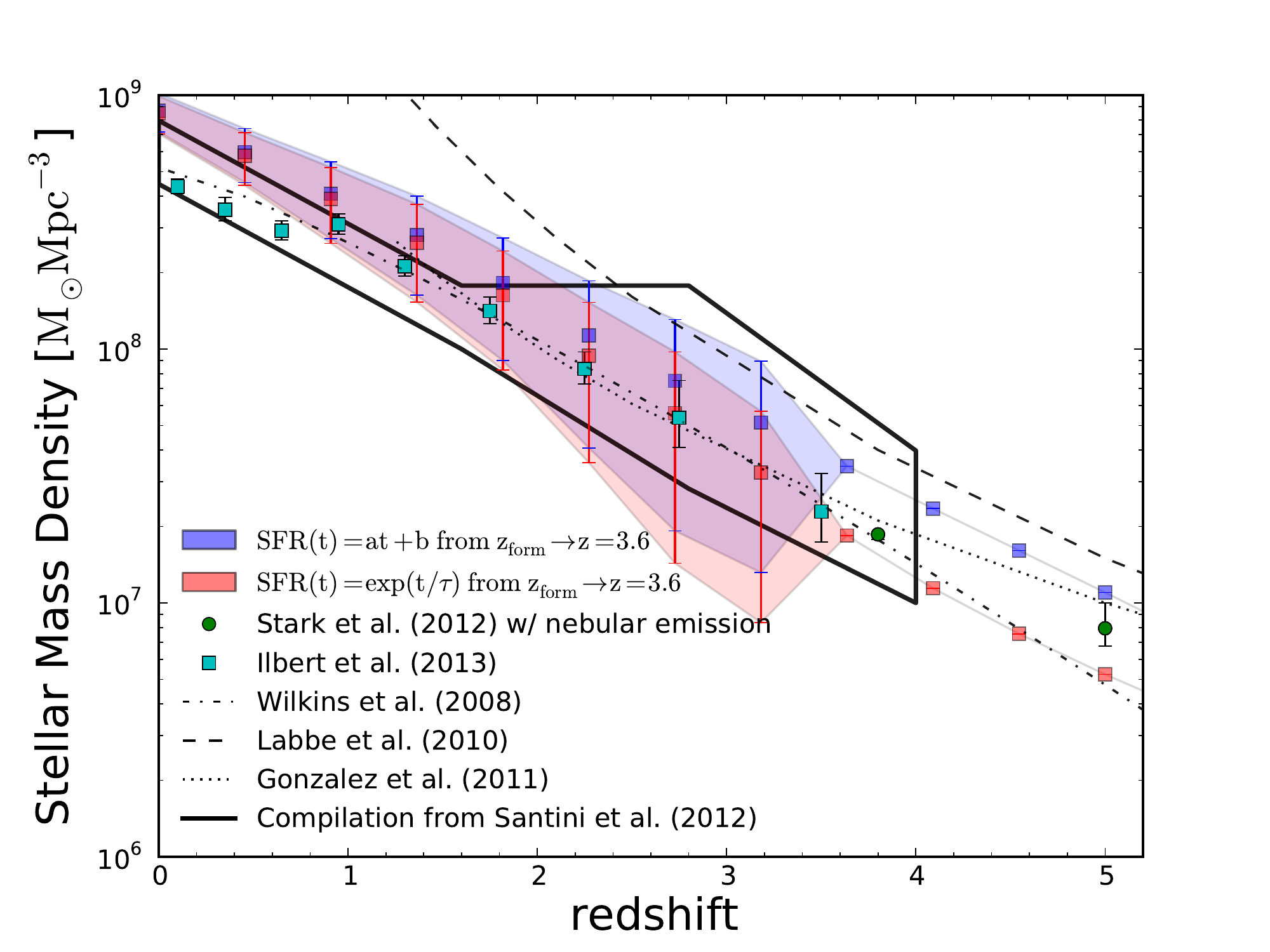}
      \caption{Evolution of the cosmic SMD vs. redshift. The lines with blue and red boxes are the mean values evaluated by integration while the lighter colors show the uncertainties from the 2 000 runs as in previous figures. We overplot the trends found by \cite{Labbe10} at $z = 7 - 8$,  \cite{Gonzalez11} and \cite{Stark13} at 3 $<$ z $<$ 7. We also plot the line corresponding to the compilation of published measurements by \cite{Wilkins08}. Within the uncertainties, we find a good agreement of the SMD integrated from the SFRD with all other SMDs based on galaxy surveys from $z = 0.6$ to $z = 5$. The black-limited area represents the compilation of results by \cite{Santini12}. Scaled to the same cosmology and IMF (Salpeter). Using \cite{Gonzalez11}, we integrate down to luminosities equivalent to $M \sim 2500 M_\odot$. 
              }
         \label{FigSMD}
   \end{figure*}

By integrating the SFRD, we can estimate the stellar-mass density (SMD) (Fig.~\ref{FigSMD} and Tab.~\ref{TabData}). To do so, we stress that we set the mass fraction of a generation of stars that is returned to the interstellar medium to a fixed value $R = 0.3$ (\cite{Fraternali12}). We also have to assume a star formation history from $z = 3.6$ up to the galaxy formation set at $z_{form} = 10$. Option 1 is a linear extrapolation while option 2 corresponds to a rising exponential $e^{t/\tau}$ with $\tau = 0.42$ as in \cite{Papovich11} that joins the observationally-deduced SFRD\footnote{Selecting option 1 or 2 does not impact on Fig.~\ref{FigSFRD}.}.

Superimposed in Fig.~\ref{FigSMD} are recent SMDs (converted to Salpeter IMF if needed). \cite{Stark13} account for the nebular emission lines contribution to the broad-band fluxes used to infer stellar masses (\cite{Ono10, deBarros12}). The trend from \cite{Labbe10} lies above our points. The others SMDs are in agreement within the uncertainties at $0.6 < z < 3.6$ for the two above options. \cite{Wilkins08} compile measurements of the SMD from the literature and provide a best fit parametric law $\rho_\star (z) = ae^{-bz^c}$ where $a = 0.0023$, $b = 0.68$ and $c = 1.2$. We also overplot it in Fig.~\ref{FigSMD}. This curve slightly underestimates our SMF at very low redshifts but at higher redshifts, it follows the points derived from our data and our assumptions.

We reach a fair agreement, especially at $0.6 < z < 3.6$. The discrepancy previously observed is reduced here but still marginally consistent at very low redshifts. As shown in Fig.~\ref{FigSFRD}, our total SFRD generally lies below \cite{Hopkins06} suggesting that this previous evaluation of dust attenuation might be over-estimated in this redshift range. Making use of FIR data allows us to reach a better agreement. Note that \cite{Hopkins06} did not directly use MIR-based data to estimate their best-fitting parametric curve but only to correct the obscuration of the FUV data.

%

\begin{figure*}
   \centering
   \includegraphics[width=\hsize]{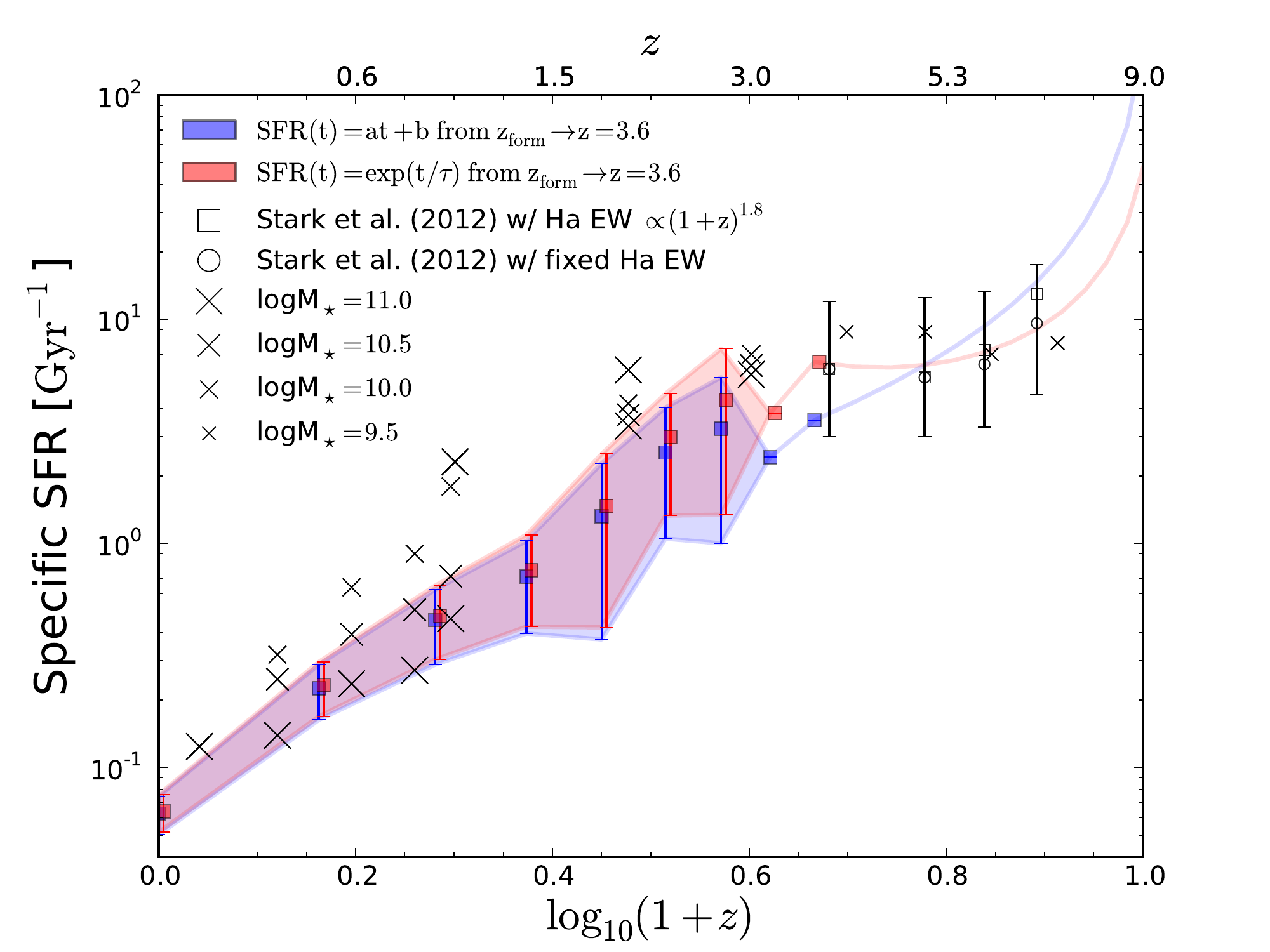}
      \caption{Cosmic sSFR vs. z extrapolated beyond the observed limit at $z = 3.6$ with different options (see text). At z < 4, crosses are extracted from several galaxy surveys and are shown as a function of the considered $\log_{10} M_\star$ by \cite{Noeske07}, \cite{Daddi07} and \cite{Wuyts11} and at $z > 4$ by \cite{Bouwens12}. Below $z = 3.6$, our data suggest that the sSFR are dominated by $\log_{10} M_\star \sim 10.5 - 11.0$ galaxies. At $z > 4$ the extrapolations are in agreement with $\log_{10} M_\star \sim 9.5$ as evaluated by \cite{Bouwens12}.
                     }
         \label{FigsSFRz}
   \end{figure*}

Fig.~\ref{FigsSFRz} shows the variation of the cosmic specific star formation rate (sSFR = SFRD/SMD) as a function of redshift. Previous observational results often suggest a flat evolution of the sSFR at $z > 2$ (\cite{Gonzalez12, Bouwens12, Schaerer13}) while most theoretical models predict a continuous rise (\cite{Bouche10, Weinmann11, Dave11}). No matter what hypothesis selected to extrapolate the observed SFRD beyond $z > 3.6$ to $z_{form} = 10$, our sSFRs remain consistent with an increase at low redshifts. The influence of the SFRD assumed at $z > 3.6$ is not noticeable within the uncertainties at $z < 3.6$. A comparison with the sSFR of galaxies from \cite{Noeske07}, \cite{Daddi07}, \cite{Wuyts11} and \cite{Bouwens12} in Fig.~\ref{FigsSFRz} suggests that the most massive galaxies ($log_{10} M_\star = 10^{10.5} - 10^{11}  [M_\odot]$) are in agreement with our sSFRs (we corrected the SMDs to $R = 0.3$ and we applied a correction to the calibration to SFR if necessary). Finally, we note that at higher redshifts, option 1 keeps on rising while option 2 shows a flattening at 4 $<$ z $<$ 5 followed by an increase at $z > 5$. We stress, though, that by assuming an exponential rise above $z > 4$ with a value of the time constant $\tau$ = 420 Myrs as in \cite{Papovich11} and the formation of galaxies at ${\rm z_{form} = 10}$ implies that the observed plateau at 4 $<$ z $<$ 5 must be temporary. Changing $\tau$ and/or ${\rm z_{form}}$ to larger redshifts would shift the increase in sSFR to earlier times. Theoretically, fixing ${\rm z_{form} = \infty}$ would translate into a flat sSFR.


\section{Discussion and Conclusions}

The variation of the cosmic dust attenuation with redshift as estimated from the IR to FUV luminosity ratio suggests the presence of a peak in the dust attenuation at $z \sim 1.2$ followed by a decline up to $z = 3.6$. This result confirms \cite{Cucciati12} dust attenuation estimated from SED fitting without FIR data. Moreover, the redshift evolution of the volume-corrected $IRX-\beta$ points globally follows the $IRX-\beta$ law from $z = 0.4$ to $z = 3.6$.

The total (FUV+FIR) cosmic SFRD starts declining above $z =  3 - 4$ and reaches the same level at $z \sim 5 - 6$ as is measured locally if we assume no variations in this trend. At $z = 3 - 4$, the decrease observed in the SFRD is not unexpected: most high redshift studies clearly suggest such a trend through rest-frame UV observations (\cite{Hopkins06, Bouwens11}) and predicted by \cite{Bethermin12b}. Backwards in time from today, this decline is preceded by a rise from $z \sim 0$ to a break at $z \sim 1 - 2$, followed by a plateau up to $z = 3 - 4$. 

If we compare Fig.~\ref{FigAfuv} to Fig.~\ref{FigSFRD}, the peak of the dust attenuation is delayed with respect to the plateau of the total SFRD. It is also coeval with the final decrease at $z = 1 - 1.5$, the peak of the less luminous AGNs (\cite{Hopkins07}). A similar peak seems to appear at $z = 1 - 2$ (depending on the stellar mass) of the cosmological merger rate (\cite{Conselice08}) and of the CIB (\cite{Bethermin12a}). Are all these effects related to the same physical phenomena and what are their characteristic timescales? To better understand the delay, it is necessary to perform an analysis via models that are fed with data of the gas content and the evolution of metallicity.

Using the observed cosmic SFRD along with the assumption of an exponential rise from $z_{form}$ = 10 to $z = 3.6$, we are able to recover the SMD evaluated from galaxy surveys. With the same assumption, we predict a flattening of the sSFR at 3 $<$ z $<$ 5 followed by a new steepening at $z > 5$.

Fig.~\ref{FigAfuv} and Fig.~\ref{FigSFRD} taken together at face value would suggest that the universe's dusty era (meaning dust attenuation larger than in the local universe) started at z = 3 - 4 simultaneously with the rise of a universe-wide star-formation event.

Fig.~\ref{FigSFRD}, Fig.~\ref{FigSMD}, and Fig.~\ref{FigsSFRz} allow us to follow the SFRD, the SMD, and the sSFR over most of the Hubble time in a consistent way. However, large uncertainties prevent us from closing the case. Additionaly, it remains quite puzzling that GRB-based analyses suggest a much shallower decrease (\cite{Kistler09, Robertson12}) than Lyman break galaxies. The statistical significance of these results is still debated because of the low number of objects at high redshifts and a possible modification of the IMF (\cite{Dwek11, Hayward13}). Another possibility is that GRBs might still be biased toward certain types of SFGs, even though this bias may be less than thought a few years ago.


\begin{acknowledgements}
      PACS has been developed by a consortium of institutes led by MPE (Germany) and including UVIE (Austria); KU Leuven, CSL, IMEC (Belgium); CEA, LAM (France); MPIA (Germany); INAF-IFSI/ OAA/OAP/OAT, LENS, SISSA (Italy); IAC (Spain). This development has been supported by the funding agencies BMVIT (Austria), ESA-PRODEX (Belgium), CEA/CNES (France), DLR (Germany), ASI/INAF (Italy), and CICYT/MCYT (Spain). SPIRE has been developed by a consortium of institutes led by Cardiff Univ. (UK) and including: Univ. Lethbridge (Canada); NAOC (China); CEA, LAM (France); IFSI, Univ. Padua (Italy); IAC (Spain); Stockholm Observatory (Sweden); Imperial College London, RAL, UCL-MSSL, UKATC, Univ. Sussex (UK); and Caltech, JPL, NHSC, Univ. Colorado (USA). This development has been supported by national funding agencies: CSA (Canada); NAOC (China); CEA, CNES, CNRS (France); ASI (Italy); MCINN (Spain); SNSB (Sweden); STFC, UKSA (UK); and NASA (USA). The authors acknowledge financial contribution from the contracts PRIN-INAF 1.06.09.05 and ASI-INAF I00507/1 and I005110. SPIRE has been developed by a consortium of institutes led by Cardiff University (UK) and including University of Lethbridge (Canada); NAOC (China); CEA, OAMP (France); IFSI, University of Padua (Italy); IAC (Spain); Stockholm Observatory (Sweden); Imperial College London, RAL, UCL-MSSL, UKATC, University of Sussex (UK); and Caltech/ JPL, IPAC, University of Colorado (USA). This development has been supported by national funding agencies: CSA (Canada); NAOC (China); CEA, CNES, CNRS (France); ASI (Italy); MCINN (Spain); Stockholm Observatory (Sweden); STFC (UK); and NASA (USA). The data presented in this paper will be released through the Herschel Database in Marseille (HeDaM; http://hedam.oamp.fr/HerMES). This work makes use of TOPCAT (http://www.star.bristol.ac.uk/?mbt/topcat/ ).
\end{acknowledgements}

 \end{document}